# A novel discrete grey seasonal model and its applications

Weijie Zhou[1], Jiao Pan[1], Song Ding[2, 3*], Xiaoli Wu[1],

1. Business college, Changzhou University, Jiangsu Changzhou 213164, China

2. School of Economics, Zhejiang University of Finance & Economics, Hangzhou 310018, China

3. Research Center for Regional Economy & Integrated Development, Zhejiang University of Finance & Economics, Hangzhou, 310018, China

**Abstract:** In order to accurately describe real systems with seasonal disturbances, which normally appear monthly or quarterly cycles, a novel discrete grey seasonal model, abbreviated as $DGSM(1,1)$, is put forward by incorporating the seasonal dummy variables into the conventional model. Moreover, the mechanism and properties of this proposed model are discussed in depth, revealing the inherent differences from the existing seasonal grey models. For validation and explanation purposes, the proposed model is implemented to describe three actual cases with monthly and quarterly seasonal fluctuations (quarterly wind power production, quarterly $PM_{10}$, and monthly natural gas consumption), in comparison with five competing models involving grey prediction models ( $SFGM(1,1)$, $SGM(1,1)$, and $DGGM(1,1)$ ), conventional econometric technology ( $SARIMA$ ), and artificial intelligences ( $BPNN$ ). Experimental results from the cases consistently demonstrated that the proposed model significantly outperforms the other benchmark models in terms of several error criteria. Moreover, further discussions about the influences of different sequence lengths on the forecasting performance reveal that the proposed model still performs the best with strong robustness and high reliability in addressing seasonal sequences. In general, the new model is validated to be a powerful and promising methodology for handling sequences with seasonal fluctuations.

**Keywords:** grey prediction; seasonal fluctuations; discrete grey seasonal model; time series modeling

---

* Corresponding author. School of Economics, Zhejiang University of Finance & Economics, Hangzhou 310018, China.
E-mail address: dings1129@zufe.edu.cn (Song DING).

# 1 Introduction

Accurate predictions can provide decision-makers with valuable information regarding the development trends of economic and social systems in the coming years. Currently, as the systems evolve quickly and are disturbed by more complex factors, accurate forecasts are confronted with various difficulties, especially for systems disturbed by seasonality and uncertainty. Generally, there are large quantities of seasonal time series in real life, such as electricity consumption [1], rainfall capacity [2], and shale gas production [3], and they can be classified into hourly, weakly, daily, monthly, quarterly, and annual sequences from the perspective of time-frequency. The prediction of seasonal data can evaluate the mid-term development trends of the time sequence to formulate relevant policy recommendations [4-5]. The seasonal data usually show some similarities between different cycle periods, and have some fluctuations and randomness [6-8], which makes accurate predictions more difficult.

Currently, methodologies of predicting seasonal time series in previous literature can be roughly sorted into three categories: statistical econometric models, artificial intelligence models, and grey models. As the representative of the statistical econometric models, the Seasonal Autoregressive Integrated Moving Average ($SARIMA$) model is often used to simulate and predict seasonal time series. By analyzing the monthly traffic data of international container ports from 1999 to 2007, Javed and Ghim [9] used the $SARIMA$ model to make reliable predictions on the throughput of major international ports. The $SARIMA$ model also successfully predicted the trends of Hong Kong's airport monthly passenger traffic from 2011 to 2015 [10]. Through the empirical study on the monthly passenger flow forecast of the Serbian railway, the advantages of the $SARIMA$ model in processing the seasonal periodicity of monthly data were demonstrated [11]. Based on the novel Hyndman-Khandakar (HK) algorithm, Gökhan [12] used the $SARIMA$ model to handle the seasonality of the Turkish inflation rate. Besides, the $SARIMA$ model can also be used to forecast quarterly data. Mwanga et al. [13] applied the $SARIMA$ model to project quarterly sugarcane yield and obtained satisfactory results. In addition, Artificial intelligence models have been widely used in the simulation and prediction of the seasonal sequence because they can describe the complex structure of the sequences. Danandeh et al. [14] used the artificial neural network ($ANN$) model to predict the monthly streamflow. Based on the ensemble $ANN$ models, Khwaja et al. [15] improved the stability and prediction accuracy in forecasting monthly power load. Li et al. [16] proposed an improved back propagation neural network ($PCA-ADE-BPNN$) model to forecast Beijing's monthly inbound passenger flow from 2011 to 2016 based on the Baidu Index. The new method exhibited higher prediction accuracy than other models. Besides, the Support vector regression ($SVR$) model can also be adapted to describe the characteristics of seasonal time series. Compared with the multiple linear regression ($MLR$) and $ANN$ models, the $SVR$ model had a smaller prediction error in forecasting the monthly natural gas consumption of Istanbul in Turkey [17]. Xie et al. [18] found that the hybrid models on the basis of $LSSVR$ had advantages in describing the seasonal and nonlinear characteristics of monthly container throughput, and possessed better forecasting performance than other models.

Statistical measurement and artificial intelligence models need a large amount of data to ensure high prediction accuracy. However, due to data availability in practical applications, a certain system sometimes has limited and incomplete data. Under such a situation, complex models are not always preferred to simpler ones in time series forecasting literature, especially confronting cases with limited and insufficient data [19]. To this end, Deng [20-21] put forward the grey prediction theory in 1982, among which various grey prediction models with simple structure and excellent performances are designed for modeling time sequences featured by uncertainty, incomplete information, and scarce data. Most importantly, the grey prediction models can achieve accurate forecasts without the dependence on large sample data or known data distributions [22-25]. Accordingly, grey models have been successfully used in many fields, such as power consumption [26], $CO_2$ emissions [27], the annual output of shale gas [28], air quality index prediction [29], and high-tech industries [30]. However, the above applications of grey models are mostly associated with annual time series. Currently, there are few studies on seasonal sequences by grey prediction models. In order to depict the fluctuant features of seasonal sequences, Wang et al. [31] built a seasonal fluctuation grey model ($SFGM(1,1)$) to predict the monthly electricity demand in South Australia. Applying a seasonal accumulating generation operator, Wang et al. [32] proposed a seasonal $GM(1,1)$ model ($SGM(1,1)$) to forecast the quarterly electricity consumption of the Chinese primary industries. The results showed that the $SGM(1,1)$ model possessed higher accuracy than its competitors. Generally, the $SFGM(1,1)$ and $SGM(1,1)$ used a similar idea of designing a seasonal factor to remove the seasonality hidden in the time sequence. Then, based on the preprocessed series, the $GM(1,1)$ models thereby are built. Afterward, by dividing a seasonal time series into different sets with the same month or quarter, Wang et al. [33] introduced a data grouping grey model $DGGM(1,1)$ to forecast the quarterly hydropower production. Experimental results demonstrated its superiority over other benchmarks.

In general, the core innovation of the $DGGM(1,1)$ model is grouping the data into different seasonal subsets, then the $GM(1,1)$ model is used to simulate and predict each subset. The profile of the $SFGM(1,1)$ and $SGM(1,1)$ models is the establishment of $GM(1,1)$ based on the sequences removed seasonality by seasonal factors. Hence, the consistent idea of the above three models is to establish the $GM(1,1)$ model based on the preprocessed data sets which have no seasonal component. Therefore, the dynamic seasonal characteristics hidden in the original data can not be effectively accounted for by the established models. Additionally, the above-modified models are only used in a given area, while more validations in diverse domains were ignored in modeling. Furthermore, numerical tests on generalization, robustness, and reliability of these above models are always absent in existing forecasting studies, not to mention a comprehensive analysis. However, according to requirements for modeling time sequences, a comprehensive analysis for testing the generalization, robustness, and reliability of a novel model is indispensable. Therefore, the main innovation of this study is to put forward a novel discrete grey seasonal model by incorporating the dummy variables into the model structure. This new model has dynamic adaptability to many time series with seasonal fluctuations, which can be validated by three different cases. Furthermore, experiments on time series with different sequence lengths are conducted to further demonstrate the generalization, robustness, and reliability of the proposed model.

The rest of this study is organized as follows. Section 2 describes the proposed methodology in detail, where mathematical derivations and properties are thoroughly discussed. Taking three cases of quarterly power production, quarterly $PM_{10}$ emissions, and monthly natural gas consumption in China as sample data, the experimental study is performed in Section 3. Moreover, the robustness of the proposed model for different sequences lengths is also tested here. Finally, Section 4 summarized this paper and presents a future research direction.

## 2 Methodology formulation

Three main subsections are introduced to the methodology applied in this study. Section 2.1 concentrates on the basic concepts of the conventional discrete grey model, e.g. $DGM(1,1)$. The elaboration of the proposed $DGSM(1,1)$ model is presented in Section 2.2. Lastly, several properties are discussed in Section 2.3 for a better understanding of the novel model.

### 2.1 Basic concepts of the DGM(1,1) model

The conventional $DGM(1,1)$ model, put forward by Xie and Liu [34], is a useful tool for confronting uncertain issues existing in insufficient information systems with sparse data. In comparison with the $GM(1,1)$ model, this model has significant advantages in reducing the inherent errors generated by the transformation from the discrete function to the continuous one. Thus, the detailed procedures of the $DGM(1,1)$ model can be outlined below:

**Step one:** Obtain the raw and transformed time series for modeling.

Assume that $X^{(0)} = (x^{(0)}(1), x^{(0)}(2), \cdots, x^{(0)}(n))$ is an original non-negative varying sequence, where $n$ is the length of the sequence. Subsequently, by using the one-order accumulation generation ($1-AGO$), these above raw data can be transformed into $X^{(1)} = (x^{(1)}(1), x^{(1)}(2), \cdots, x^{(1)}(n))$, for which the kth entry is denoted as $x^{(1)}(k) = \sum_{i=1}^{k} x^{(0)}(i)$, $k = 1, 2, \cdots, n$.

**Step two:** Build the conventional discrete grey model.

$$x^{(1)}(k+1) = \beta_1 x^{(1)}(k) + \beta_2 \tag{1}$$

is called the discrete grey model, abbreviated as $DGM(1,1)$. In this model, $\beta_1$ is the development coefficient of the system and $\beta_2$ is a grey constant.

**Step three:** Estimate the model parameters.

Substitute the values of $k$ into Eq. (1), we can obtain

$$x^{(1)}(2)=\hat{\beta}_1 x^{(1)}(1)+\hat{\beta}_2$$
$$x^{(1)}(3)=\hat{\beta}_1 x^{(1)}(2)+\hat{\beta}_2$$
$$\vdots$$
$$x^{(1)}(n)=\hat{\beta}_1 x^{(1)}(n-1)+\hat{\beta}_2 \qquad (2)$$

In the matrix form, $Y = CD$, where

$$C = \begin{bmatrix} x^{(1)}(1) & 1 \\ x^{(1)}(2) & 1 \\ \vdots & \vdots \\ x^{(1)}(n-1) & 1 \end{bmatrix}, Y = \begin{bmatrix} x^{(1)}(2) \\ x^{(1)}(3) \\ \vdots \\ x^{(1)}(n) \end{bmatrix}, D = \begin{bmatrix} \hat{\beta}_1 \\ \hat{\beta}_2 \end{bmatrix}. \qquad (3)$$

Solving the matrix form of the equation by using the least square estimation, the parameters can be achieved as $D = \left[\hat{\beta}_1, \hat{\beta}_2\right]^T = (C^T C)^{-1} C^T Y$.

**Step four:** Calculate the time response function for generating predictions in the transformed domain.

After estimating the parameters and using $x_1^{(1)}(1) = x_1^{(0)}(0)$ as the initial condition, the time response function can be achieved as the following formula,

$$x^{(1)}(k+1)=\hat{\beta}_1^k x^{(0)}(1)+\frac{1-\hat{\beta}_1^k}{1-\hat{\beta}_1}\bullet\hat{\beta}_2, \, k=1,2,\cdots. \qquad (4)$$

Substituting the values of $k$, the fitted and predicted values in the transformed domain can be obtained.

**Step five:** Obtain the restored values in the original domain.

By using the first-order inversed accumulating generation operator (*1-IAGO*), the fitted and predicted values in the original domain can be calculated by

$$\hat{x}^{(0)}(k+1) = (\hat{\beta}_1 - 1)(x^{(0)}(1) - \frac{\hat{\beta}_2}{1-\hat{\beta}_1})\hat{\beta}_1^k, k=1,2,\cdots \qquad (5)$$

where, $\hat{x}^{(1)}(k)(k=1,2,\cdots,n)$ are called fitted values, and $\hat{x}^{(0)}(k)(k \geq n+1)$ are called predicted values.

By using these above five steps, the traditional discrete grey model can be successfully adapted to various areas. Evidence from previous studies show that this model can significantly improve forecasting performance, especially for modeling the pure index sequences [34]. Thus, this discrete model performs better than the $GM(1,1)$ model theoretically. However, confronting with some nonlinear sequences, such as seasonal series, which are practically ubiquitous, this model may be ineffective in generating accurate forecasts. Therefore, designing a novel model for handing such seasonal series is imperative for researchers and practitioners.

**2.2 The proposed DGSM(1,1) model for modeling seasonal sequences**

Considering the inherent drawbacks in the conventional discrete grey model, we put forward a novel grey model for modeling seasonal sequences. The principle of this model is illustrated below.

**Definition 1.** Assume that $X^{(0)}$ and $X^{(1)}$ are the same as defined in Section 2.1, and

$$x^{(1)}(k+1) = \alpha x^{(1)}(k) + \beta_{M(k+1,s)}, \quad k = 1, 2, \cdots, \qquad (6)$$

is called as a novel Discrete Grey Seasonal Model having one variable and one order, abbreviated as $DGSM(1,1)$, where $s$ is the number of the seasonal cycle, $\beta_{M(k+1,s)}$ represents the seasonal item, and

$$M(k,s) = \begin{cases} s & k \bmod s = 0 \\ k \bmod s & k \bmod s \neq 0 \end{cases}. \qquad (7)$$

The first "1" in the $DGSM(1,1)$ model represents only one variable, while the second "1" is the first order accumulation operation, which can significantly reduce the randomness hidden in the original data set. For instance, $s = 4$ *and* $12$ stand for the quarterly and monthly sequences, respectively, thereby the corresponding parameters in Eq. (6) are remarked as $(\alpha, \beta_1, \cdots, \beta_4)$ and $(\alpha, \beta_1, \cdots, \beta_{12})$.

**Theorem 1.** Set that $P = (\alpha, \beta_1, \cdots, \beta_s)$ is the parameter vector of the $DGSM(1,1)$ model. Using the ordinary least square (OLS) method, the estimated parameter vector $\hat{P}$ can be calculated as $\hat{\mathbf{P}} = (\hat{\alpha}, \hat{\beta}_1, \cdots, \hat{\beta}_s) = (\mathbf{B}^{\mathrm{T}}\mathbf{B})^{-1} \mathbf{B}^{\mathrm{T}}\mathbf{Y}$, where

$$\mathbf{B} = \begin{bmatrix} x^{(1)}(1) & 0 & 1 & 0 & \cdots & \cdots & \cdots & 0 \\ x^{(1)}(2) & 0 & 0 & 1 & \cdots & \cdots & \cdots & 0 \\ \vdots & \vdots & \vdots & \vdots & \vdots & \vdots & \vdots & \vdots \\ x^{(1)}(s-1) & 0 & 0 & 0 & \cdots & \cdots & \cdots & 1 \\ x^{(1)}(s) & 1 & 0 & 0 & \vdots & \vdots & \vdots & 0 \\ \vdots & \vdots & \vdots & \vdots & \vdots & \vdots & \vdots & \vdots \\ x^{(1)}(n-1) & \cdots & \cdots & \cdots & \cdots & 1 & \cdots & 0 \end{bmatrix}, \quad \mathbf{Y} = \begin{bmatrix} x^{(1)}(2) \\ x^{(1)}(3) \\ \vdots \\ x^{(1)}(n) \end{bmatrix},$$

**Proof.** Substituting $x^{(1)}(k)$ into Eq. (6), we get

$$\begin{cases} x^{(1)}(2) = \hat{\alpha} x^{(1)}(1) + 0 \cdot \hat{\beta}_1 + 1 \cdot \hat{\beta}_2 + 0 \cdot \hat{\beta}_3 + \cdots + 0 \cdot \hat{\beta}_s \\ x^{(1)}(3) = \hat{\alpha} x^{(1)}(2) + 0 \cdot \hat{\beta}_1 + 0 \cdot \hat{\beta}_2 + 1 \cdot \hat{\beta}_3 + \cdots + 0 \cdot \hat{\beta}_s \\ \vdots \\ x^{(1)}(s) = \hat{\alpha} x^{(1)}(s-1) + 0 \cdot \hat{\beta}_1 + 0 \cdot \hat{\beta}_2 + 0 \cdot \hat{\beta}_3 + \cdots + 1 \cdot \hat{\beta}_s \\ x^{(1)}(s+1) = \hat{\alpha} x^{(1)}(s) + 1 \cdot \hat{\beta}_1 + 0 \cdot \hat{\beta}_2 + 0 \cdot \hat{\beta}_3 + \cdots + 0 \cdot \hat{\beta}_s \\ \vdots \\ x^{(1)}(n) = \hat{\alpha} x^{(1)}(n-1) + \cdots + 1 \cdot \hat{\beta}_{M(n,s)} + \cdots + 0 \cdot \hat{\beta}_s \end{cases} \quad (8)$$

Transforming into the matrix form, Eq. (8) can be equivalent to $Y = BP$. Based on the Ordinary Least Squares (OLS) method, the estimated parameter vector $\hat{P}$ is given as $\hat{\mathbf{P}} = (\mathbf{B}^T \mathbf{B})^{-1} \mathbf{B}^T \mathbf{Y}$. Once we have the estimated values of all the parameters, the forecasting function in Theorem 2 will be obtained for future estimations.

**Theorem 2.** Let that $\hat{\mathbf{P}} = (\hat{\alpha}, \hat{\beta}_1, \cdots, \hat{\beta}_s)$ is the estimated parameter vector of the $DGSM(1,1)$ model, the time response function for predicting seasonal sequences in the transformed domain are provided as follows:

$$\hat{x}^{(1)}(k) = \hat{\alpha}^{k-1} x^{(0)}(1) + \hat{\alpha}^{k-2} \hat{\beta}_2 + \sum_{j=3}^{k} \hat{\alpha}^{k-j} \hat{\beta}_{M(j,s)}, \quad (9)$$

and its corresponding restored function in the original domain is

$$\hat{x}^{(0)}(k) = \hat{\alpha}^{k-2}(\hat{\alpha} - 1) x^{(0)}(1) + \hat{\alpha}^{k-2} \hat{\beta}_2 + \sum_{j=3}^{k} \hat{\alpha}^{k-j} (\hat{\beta}_{M(j,s)} - \hat{\beta}_{M(j-1,s)}), \quad k \geq 2 \quad (10)$$

where $\hat{x}^{(1)}(1) = \hat{x}^{(0)}(1) = x^{(0)}(1)$.

**Proof.** The mathematical induction method is adopted to prove Theorem 2.

When $k = 2$, based on Eq. (6), we have

$$\hat{x}^{(1)}(2) = \hat{\alpha}\hat{x}^{(1)}(1) + \hat{\beta}_2 = \hat{\alpha}x^{(0)}(1) + \hat{\beta}_2 = \hat{\alpha}^{2-1}x^{(0)}(1) + \hat{\alpha}^{2-2}\hat{\beta}_2 + \sum_{j=3}^{2}\hat{\alpha}^{2-j}\hat{\beta}_{M(j,s)} \quad (11)$$

Assuming Eq. (9) is correct when $k = m$, then we can obtain the following equation when $k = m+1$:

$$\begin{aligned}
\hat{x}^{(1)}(m+1) &= \hat{\alpha}\hat{x}^{(1)}(m) + \hat{\beta}_{M(m+1,s)} \\
&= \hat{\alpha}\left(\hat{\alpha}^{m-1}x^{(0)}(1) + \hat{\alpha}^{m-2}\hat{\beta}_2 + \sum_{j=3}^{m}\hat{\alpha}^{m-j}\hat{\beta}_{M(j,s)}\right) + \hat{\beta}_{M(m+1,s)} \\
&= \hat{\alpha}^{m+1-1}x^{(0)}(1) + \hat{\alpha}^{m+1-2}\hat{\beta}_2 + \sum_{j=3}^{m}\hat{\alpha}^{m+1-j}\hat{\beta}_{M(j,s)} + \hat{\alpha}^{m+1-m-1}\hat{\beta}_{M(m+1,s)} \\
&= \hat{\alpha}^{m+1-1}x^{(0)}(1) + \hat{\alpha}^{m+1-2}\hat{\beta}_2 + \sum_{j=3}^{m+1}\hat{\alpha}^{m+1-j}\hat{\beta}_{M(j,s)}
\end{aligned} \quad (12)$$

Obviously, Eq. (9) is also correct when $k = m+1$, demonstrating that the time response function in the transformed domain is proved to be right. Subsequently, by using the inverse transformation $1-IAGO$, we achieve

$$\begin{aligned}
x^{(0)}(k) &= \hat{x}^{(1)}(k) - \hat{x}^{(1)}(k-1) \\
&= \hat{\alpha}^{k-2}x^{(0)}(1) + \hat{\alpha}^{k-2}\hat{\beta}_2 + \sum_{j=3}^{k}\hat{\alpha}^{k-j}\hat{\beta}_{M(j,s)} - [\hat{\alpha}^{k-2}x^{(0)}(1) + \hat{\alpha}^{k-3}\hat{\beta}_2 + \sum_{j=3}^{k-1}\hat{\alpha}^{k-1-j}\hat{\beta}_{M(j,s)}] \\
&= \hat{\alpha}^{k-2}(\hat{\alpha}-1)x^{(0)}(1) + \hat{\alpha}^{k-2}\hat{\beta}_2 + \hat{\alpha}^{k-3}\hat{\beta}_{M(3,s)} + \sum_{j=4}^{k}\hat{\alpha}^{k-j}\hat{\beta}_{M(j,s)} - \hat{\alpha}^{k-3}\hat{\beta}_2 - \sum_{j=3}^{k-1}\hat{\alpha}^{k-1-j}\hat{\beta}_{M(j,s)} \\
&= \hat{\alpha}^{k-2}(\hat{\alpha}-1)x^{(0)}(1) + \hat{\alpha}^{k-2}\hat{\beta}_2 + \hat{\alpha}^{k-3}(\hat{\beta}_{M(3,s)} - \hat{\beta}_{M(2,s)}) + \sum_{j=4}^{k}\hat{\alpha}^{k-j}\hat{\beta}_{M(j,s)} - \sum_{j=4}^{k}\hat{\alpha}^{k-j}\hat{\beta}_{M(j-1,s)} \\
&= \hat{\alpha}^{k-2}(\hat{\alpha}-1)x^{(0)}(1) + \hat{\alpha}^{k-2}\hat{\beta}_2 + \sum_{j=3}^{k}\hat{\alpha}^{k-j}(\hat{\beta}_{M(j,s)} - \hat{\beta}_{M(j-1,s)})
\end{aligned}$$

(13)

Thus, the second conclusion of Eq. (10) can be easily proved. In conclusion, Theorem 2 can be used for projections in the original and transformed domain.

**2.3 The properties of the DGSM(1,1) model**

From the modeling mechanism of the conventional and novel discrete grey models in Sections 2.1 and 2.2, respectively, we find that several properties of the proposed model can be concluded for better understanding its originality and extensibility.

**Property 1** If $s=1$, then the $DGSM(1,1)$ model degenerates into the conventional discrete grey model ($DGM(1,1)$).

**Proof:** if $s=1$, the collected data represents the sequences without seasonality. Then, the $\beta_{M(k+1,s)}$ in the proposed model will become a constant, meaning that the novel model is equivalent to the traditional one. In other words, the new model can unify the traditional one, thereby further extending the application areas.

**Property 2** The novel model is unbiased to the seasonal sequences. To be specific, assuming the first-stage series $\hat{X}^{(0)} = (\hat{x}^{(0)}(1), \hat{x}^{(0)}(2), \cdots, \hat{x}^{(0)}(n))$ is generated by using the proposed model based on the collected seasonal sequence $X^{(0)} = (x^{(0)}(1), x^{(0)}(2), \cdots, x^{(0)}(n))$, then the second-stage series $\hat{\hat{X}}^{(0)} = (\hat{\hat{x}}^{(0)}(1), \hat{\hat{x}}^{(0)}(2), \cdots, \hat{\hat{x}}^{(0)}(n))$ will be calculated on the basis of the first-stage sequence. Compared with these two sequences for the first and second stages, we find that they are equivalent, namely $\hat{\hat{x}}^{(0)}(k) = \hat{x}^{(0)}(k), k = 1, 2, \cdots, n$.

**Proof:** Through the reverse process of Theorem 2, Property 2 can be proved, meaning the novel model is unbiased for seasonal sequences. The mathematics certification was deduced in detail, which is presented below.

For the first stage, after establishing the proposed model based on the original series $X^{(0)} = (x^{(0)}(1), x^{(0)}(2), \cdots, x^{(0)}(n))$, we can obtain the forecasted sequence in the original domain, which is marked as

$$\hat{x}^{(0)}(k) = \hat{\alpha}^{k-2}(\hat{\alpha}-1)x^{(0)}(1) + \hat{\alpha}^{k-2}\hat{\beta}_2 + \sum_{j=3}^{k} \hat{\alpha}^{k-j}\left(\hat{\beta}_{M(j,s)} - \hat{\beta}_{M(j-1,s)}\right) \tag{14}$$

Subsequently, by using the 1-AGO, the accumulated sequence is

$$\begin{aligned}\hat{x}^{(1)}(k)&=\sum_{i=1}^{k}\hat{x}^{(0)}(i)=\hat{x}^{(0)}(1)+\hat{x}^{(0)}(2)+\hat{x}^{(0)}(3)+\cdots+\hat{x}^{(0)}(k-1)+\hat{x}^{(0)}(k)\\ &=x^{(0)}(1)+(\hat{\alpha}-1)x^{(0)}(1)+\hat{\beta}_2+x^{(0)}(1)+\hat{\alpha}(\hat{\alpha}-1)x^{(0)}(1)+\hat{\alpha}\hat{\beta}_2+(\hat{\beta}_3-\hat{\beta}_2)+\cdots\\ &\quad+\hat{\alpha}^{k-1-2}(\hat{\alpha}-1)x^{(0)}(1)+\hat{\alpha}^{k-1-2}\hat{\beta}_2+\sum_{j=3}^{k-1}\hat{\alpha}^{k-1-j}(\hat{\beta}_{M(j,s)}-\hat{\beta}_{M(j-1,s)})\\ &\quad+\hat{\alpha}^{k-2}(\hat{\alpha}-1)x^{(0)}(1)+\hat{\alpha}^{k-2}\hat{\beta}_2+\sum_{j=3}^{k}\hat{\alpha}^{k-j}(\hat{\beta}_{M(j,s)}-\hat{\beta}_{M(j-1,s)})\\ &=\hat{\alpha}^{k-1}x^{(0)}(1)+\hat{\alpha}^{k-2}\hat{\beta}_2+\sum_{j=3}^{k}\hat{\alpha}^{k-j}\hat{\beta}_{M(j,s)}\end{aligned} \quad (15)$$

Similarly, we can get $\hat{x}^{(1)}(k+1)=\hat{\alpha}^{k}(\hat{\alpha}-1)x^{(0)}(1)+\hat{\alpha}^{k-1}\hat{\beta}_2+\sum_{j=3}^{k+1}\hat{\alpha}^{k+1-j}\hat{\beta}_{M(j,s)}$, thereby

$$\hat{x}^{(1)}(k+1)=\hat{\alpha}\hat{x}^{(1)}(k)+\hat{\beta}_{M(k+1,s)}.$$

For the second stage, based on the predicted values in the first stage, namely $\hat{X}^{(0)}=(\hat{x}^{(0)}(1),\hat{x}^{(0)}(2),\cdots,\hat{x}^{(0)}(n))$, we rebuild the novel model and the corresponding forecasted values can be expressed below according to Theorem 2:

$$\hat{\hat{x}}^{(1)}(k)=\hat{\alpha}^{k-1}x^{(0)}(1)+\hat{\alpha}^{k-2}\hat{\beta}_2+\sum_{j=3}^{k}\hat{\alpha}^{k-j}\hat{\beta}_{M(j,s)} \quad (16)$$

$$\hat{\hat{x}}^{(0)}(k)=\hat{\alpha}^{k-2}(\hat{\alpha}-1)x^{(0)}(1)+\hat{\alpha}^{k-2}\hat{\beta}_2+\sum_{j=3}^{k}\hat{\alpha}^{k-j}\left(\hat{\beta}_{M(j,s)}-\hat{\beta}_{M(j-1,s)}\right), k\geq 2 \quad (17)$$

Comparing the forecasted results in the first and second stages, we can conclude that $\hat{\hat{x}}^{(0)}(k)=\hat{x}^{(0)}(k), k=1,2,\cdots,n$. In other words, property 2 is proved and the proposed model is an unbiased one.

**Property 3** For an original sequence $X^{(0)}=(x^{(0)}(1),x^{(0)}(2),\cdots,x^{(0)}(n))$ that is characterized by $x^{(0)}(k+s)=x^{(0)}(k)=C_{M(k,s)}$. It describes a situation that data at a certain point remains the same for different cycles while data in the other points varies. Under such a situation, the proposed model can still make error-free forecasts with the expressions of $\hat{\alpha}=1$, $\hat{\beta}_i=C_{M(i,s)}=C_i$, $i=1,2,\cdots,s$, and $\hat{x}^{(0)}(k)=x^{(0)}(k)$.

**Proof.** Due to $x^{(0)}(k+s) = x^{(0)}(k) = C_{M(k,s)}$, we can obtain the sequences noted as $X^{(0)} = (C_1, C_2, \cdots, C_s, C_1, C_2, \cdots, C_s, \cdots, C_1, C_2, \cdots, C_{M(n,s)})$, and

$$\begin{cases} x^{(1)}(k) = \frac{k}{s}\sum_{i=1}^{s} C_i, \ x^{(1)}(k+1) = \frac{k}{s}\sum_{i=1}^{s} C_i + C_1 & k \bmod s = 0 \\ x^{(1)}(k) = \frac{(k-M(k,s))}{s}\sum_{i=1}^{s} C_i + \sum_{i=1}^{M(k,s)} C_i, \ x^{(1)}(k+1) = \frac{(k+1-M(k+1,s))}{s}\sum_{i=1}^{s} C_i + \sum_{i=1}^{M(k+1,s)} C_i & k \bmod s \neq 0 \end{cases} \quad (18)$$

When $k \bmod s = 0$, we will have $x^{(1)}(k+1) = x^{(1)}(k) + C_1 = x^{(1)}(k) + C_{M(k+1,s)}$. Conversely, when $k \bmod s \neq 0$ and $M(k+1,s) = M(k,s)+1$, we can get $x^{(1)}(k+1) = x^{(1)}(k) + C_{M(k+1,s)}$, according to Eq. (16). As $\{C_{M(k+1,s)}\}$ is a cycle sequence with $s$ periods, we only need to analyze the $x^{(1)}(k+1) = x^{(1)}(k) + C_{M(k+1,s)}$ in one cycle. Let $k = 1, 2, \cdots, s$, we can get $\hat{\alpha} = 1, \hat{\beta}_i = C_{M(i,s)} = C_i$ according to Theorem 1. Based on Theorem 2, we can get $\hat{x}^{(0)}(k) = x^{(0)}(k)$. Thus, this property is proved and illustrates that the new model performs high reliability and flexibility for addressing issues of various seasonal characteristics.

**Property 4** For a collected sequence $X^{(0)} = (x^{(0)}(1), x^{(0)}(2), \cdots, x^{(0)}(n))$, we get the simulated or forecasted values $\hat{x}^{(0)}(k)$ by establishing the novel model, then

$$\hat{x}^{(0)}(k+s) = \hat{\alpha}^s x^{(0)}(k) + \psi_{M(k,s)},$$

or $\hat{x}^{(0)}(k+s) = \hat{\alpha}^s \left( x^{(0)}(k) - \frac{\psi_{M(k,s)}}{1-\hat{\alpha}^s} \right) + \frac{\psi_{M(k,s)}}{1-\hat{\alpha}^s}$, $k \geq 2$,

where 
$$\begin{cases} \psi_1 = (1-\hat{\alpha}^{s-1})\hat{\beta}_1 + (\hat{\alpha}^{s-1}-\hat{\alpha}^{s-2})\hat{\beta}_2 + \cdots + (\hat{\alpha}^2 - \hat{\alpha})\hat{\beta}_{s-1} + (\hat{\alpha}-1)\hat{\beta}_s \\ \psi_2 = (\hat{\alpha}-1)\hat{\beta}_1 + (1-\hat{\alpha}^{s-1})\hat{\beta}_2 + \cdots + (\hat{\alpha}^3 - \hat{\alpha}^2)\hat{\beta}_{s-1} + (\hat{\alpha}^2 - \hat{\alpha})\hat{\beta}_s \\ \vdots \\ \psi_s = (\hat{\alpha}^{s-1}-\hat{\alpha}^{s-2})\hat{\beta}_1 + (\hat{\alpha}^{s-2}-\hat{\alpha}^{s-3})\hat{\beta}_2 + \cdots + (\hat{\alpha}-1)\hat{\beta}_{s-1} + (1-\hat{\alpha}^{s-1})\hat{\beta}_s \end{cases}.$$

**Proof.** According to Theorem 2, we can obtain

$$\hat{x}^{(0)}(k) = \hat{\alpha}^{k-2}(\hat{\alpha}-1)x^{(0)}(1) + \hat{\alpha}^{k-2}\hat{\beta}_2 + \sum_{j=3}^{k} \hat{\alpha}^{k-j}\left(\hat{\beta}_{M(j,s)} - \hat{\beta}_{M(j-1,s)}\right)$$

and $\hat{x}^{(0)}(k+s) = \hat{\alpha}^{k+s-2}(\hat{\alpha}-1)x^{(0)}(1) + \hat{\alpha}^{k+s-2}\hat{\beta}_2 + \sum_{j=3}^{k+s} \hat{\alpha}^{k+s-j}\left(\hat{\beta}_{M(j,s)} - \hat{\beta}_{M(j-1,s)}\right)$.

Simplifying, we can get this:

$$\hat{\alpha}^s \hat{x}^{(0)}(k) = \hat{\alpha}^s \hat{\alpha}^{k-2}(\hat{\alpha}-1)x^{(0)}(1) + \hat{\alpha}^s \hat{\alpha}^{k-2}\hat{\beta}_2 + \sum_{j=3}^{k} \hat{\alpha}^s \hat{\alpha}^{k-j}\left(\hat{\beta}_{M(j,s)} - \hat{\beta}_{M(j-1,s)}\right) \quad (19)$$

and

$$\hat{x}^{(0)}(k+s) = \hat{\alpha}^s \hat{x}^{(0)}(k) + \sum_{j=k+1}^{k+s} \hat{\alpha}^{k+s-j}\left(\hat{\beta}_{M(j,s)} - \hat{\beta}_{M(j-1,s)}\right) \quad (20)$$

As Eq.(20) is a cycle equation for the $\left\{\hat{x}^{(0)}(k)\right\}$ sequence with $s$, we will take one cycle for explanation purposes. When $k=2$, we will get

$$\begin{aligned}
\psi_2 = \psi_{M(2,s)} &= \sum_{j=2+1}^{2+s} \hat{\alpha}^{2+s-j}\left(\hat{\beta}_{M(j,s)} - \hat{\beta}_{M(j-1,s)}\right) \\
&= \hat{\alpha}^{s-1}\left(\hat{\beta}_3 - \hat{\beta}_2\right) + \hat{\alpha}^{s-2}\left(\hat{\beta}_4 - \hat{\beta}_3\right) + \cdots + \hat{\alpha}\left(\hat{\beta}_1 - \hat{\beta}_s\right) + \hat{\beta}_2 - \hat{\beta}_1 \\
&= (\hat{\alpha}-1)\hat{\beta}_1 + (1-\hat{\alpha}^{s-1})\hat{\beta}_2 + \cdots + (\hat{\alpha}^3 - \hat{\alpha}^2)\hat{\beta}_{s-1} + (\hat{\alpha}^2 - \hat{\alpha})\hat{\beta}_s
\end{aligned} \quad (21)$$

Similarly, we can obtain others $\psi_{M(k,s)}$ for $k = 3, \cdots, s, s+1$.

According to Property 4, if $\hat{\alpha} > 1$, the simulated or forecasted sequences from the $DGSM(1,1)$ model possess an upward trend, and if $\hat{\alpha} < 1$, the simulated or forecasted sequences exhibit a downward trend. Therefore, the $DGSM(1,1)$ model can be used to a seasonal sequence with an upward or downward trend.

## 3 Verifications of the DGSM(1,1) model

For competition and validation purposes, the novel model is performed to predict three different seasonal sequences involving quarterly wind power production, quarterly $PM_{10}$ prediction, and monthly natural gas consumption, in comparison with five prevalent benchmark models. For these three cases, the available sample data are collected from Wind Database (http://www.wind.com.cn/), as seen in Table 1, and generally divided into two subsets, namely the training and testing subsets. In addition, these competing models include three categories, i.e., grey prediction models ($SFGM(1,1)$ [31], $SGM(1,1)$ [32], and $DGGM(1,1)$ [33], conventional econometric technology ($SARIMA$ [9, 13]), and artificial intelligence ($BPNN$ [16]). Moreover, three main criteria are employed for measuring level prediction accuracy, namely absolute percent error ($APE$), mean absolute percent error for simulation ($MAPES$) and prediction ($MAPEP$), and root mean squared error for simulation (*RMSES*) and prediction (*RMSEP*), which are calculated by Equations (22)-(24).

$$\text{APE}(k) = \left| \frac{\hat{x}^{(0)}(k) - x^{(0)}(k)}{x^{(0)}(k)} \right| \times 100\%, \quad k = 1, 2, \cdots, n+h \quad (22)$$

$$\text{MAPES} = \frac{1}{n}\sum_{k=1}^{n} \text{APE}(k), \quad \text{MAPEP} = \frac{1}{h}\sum_{k=n+1}^{n+h} \text{APE}(k) \quad (23)$$

$$\text{RMSES} = \sqrt{\frac{1}{n}\sum_{k=1}^{n}\left(\hat{x}^{(0)}(k) - x^{(0)}(k)\right)^2}, \quad \text{RMSEP} = \sqrt{\frac{1}{h}\sum_{k=n+1}^{n+h}\left(\hat{x}^{(0)}(k) - x^{(0)}(k)\right)^2} \quad (24)$$

where $n$ represents the number of observations in the fitted dataset and $h$ is the forecasted horizon.

Table 1 Data sets used for validations

| NO. | Cases | Period | Training subset | Testing subset |
|---|---|---|---|---|
| 1 | Quarterly wind power production | 2011. Q1-2018. Q4 | 20 | 12 |
| 2 | Monthly industrial electricity consumption | 2014. Q1-2019. Q4 | 16 | 8 |
| 3 | Monthly natural gas consumption | 2009. M1-2018. M12 | 108 | 12 |

**3.1 Quarterly wind power production**

Wind power is a clean and renewable energy source, which plays a crucial role in optimizing the energy structure. Therefore, it is of great significance to accurately predict wind power generation. In this case, quarterly wind power production data in China covers the period from the first quarter in 2011 to the fourth quarter in 2018 with a total of 32 observations. For verification purpose, we select data from the first quarter in 2011 to the fourth quarter in 2015 as training subset for model calibration and testing fitting performance, while the remaining data from the first quarter in 2016 to the fourth quarter in 2018 are used as testing data for evaluating predicting performance of diverse models.

For the parameter estimations of these competitors, all results are recorded in Table 2. As observed in this table, the estimated values of the parameters in the proposed models are $\hat{\mathbf{P}} = (\hat{\alpha}, \hat{\beta}_1, \cdots, \hat{\beta}_s) = (1.05, 180.36, 213.96, 125.70, 205.50)$. According to Property 4, future values of quarterly wind power keep an upward trend due to $\hat{\alpha} > 1$, which can also be validated in Fig. 1. For the other grey models, there estimated values of parameters are obtained by using the least square method. For the $SARIMA(p,d,q)(P,D,Q)$, the best $SARIMA$ for the wind power data is determined, based on autocorrelation function (ACF), partial autocorrelation function (PACF) and the minimization of Akaike Information Criterion (AIC) [35], and the optimal form is finally specified as $SARIMA(3,0,0)(0,1,0)$. In addition, the $BPNN$ model is established in this study, using one hidden layer and seven neurons. This model is iteratively run 1000 times for model elaboration with the learning rate and error goal of 0.01 and 0.05, respectively, utilizing the training subset mentioned above.

Table 2 The estimated parameters of six competing models for quarterly wind power production

| Models | Parameters |
|---|---|
| $DGSM(1,1)$ | $\hat{P} = [1.05, 180.36, 213.96, 125.70, 205.50]$. |
| $SFGM(1,1)$ | $\hat{a} = 0.0454, \hat{b} = 179.77, I_1 = 0.9132, I_2 = 1.093, I_3 = 0.842, I_4 = 1.1518$ |
| $SGM(1,1)$ | $\hat{\alpha}_s = [-0.0454, 179.84], f_s(1) = 0.919, f_s(2) = 1.089, f_s(3) = 0.8389, f_s(4) = 1.153$. |
| $DGGM(1,1)$ | $\hat{\alpha}_1 = [-0.2150, 148.83], \hat{\alpha}_2 = [-0.1994, 181.43], \hat{\alpha}_3 = [-0.1401, 1779.61], \hat{\alpha}_4 = [-0.1543, 238.66]$ |
| $SARIMA$ | SARIMA(3,0,0)(0,1,0)$_4$, c=57.8635, AR(1)=0.3743, AR(2)=0, AR(3)=-0.6183, AIC=9.116, Log L=-68.92. |
| $BPNN$ | Learning rate=0.01, iterative number=1000, one hidden layer, error goal=0.05, the number of neurons =7. |

Therefore, in this study, six competing models, including $DGSM(1,1)$, $SFGM(1,1)$, $SGM(1,1)$, $DGGM(1,1)$, $SARIMA$, and $BPNN$ are performed to forecast quarterly wind power production in China. Fig. 1 and Table 3 illustrate the comparisons of the performance of all competing models, in terms of three statistical indexes. From this figure and table, it can be consistently concluded that the novel $DGSM(1,1)$ model is superior to all other competitors for its accurate fitting and predicting performance. To be specific, from the perspective of overall findings presented graphically in figures, Fig. 1 presents the split sample experiments for six competing models, illustrating that the proposed model obtains better forecasts because the forecasted values are much closer to the actual ones in both fitted and predicted period. It worth mentioning that although the $SARIMA$ and $BPNN$ models perform well in the fitted period, their values diverge far from the observations in the predicted period, especially for the last several data points. It means that these two non-grey models are not appropriate for future estimations due to their weak forecasting capabilities.

In addition, from the perspective of accuracy levels, results of $MAPE$ and $RMSE$ criteria support the following conclusions: the $DGGM(1,1)$ model obtains the smallest $MAPES$ and $RMSES$ (3.07% and 11.89, respectively) in the fitted period, a little smaller than the corresponding counterpart of the proposed model (6.02% and 20.52, respectively). However, in the predicted period, the novel model is the best as a result of possessing the smallest $MAPEP$ and $RMSEP$ of 9.03% and 82.45, respectively, among all competitors, showing an excellent forecasting level. By contrast, the $MAPEP$ values of all other models are all over 10%, showing good forecasting level [36-37]. In addition, for the $APE$ criteria, the values of the proposed model vary in a small range with [0.32%, 22.74%] for the predicted period, illustrating its reliable capabilities for providing accurate projections. As for the other models, wider fluctuations result in the poorer forecasting performances of each benchmark model, which are highlighted in bold in Table 3.

In general, concluded from the perspectives of graphical discussions and accuracy levels in terms of $APE$, $MAPE$ and $RMSE$ criteria, the proposed $DGSM(1,1)$ model performs the overall best among all competing models for predicting China's wind power production characterized by seasonality.

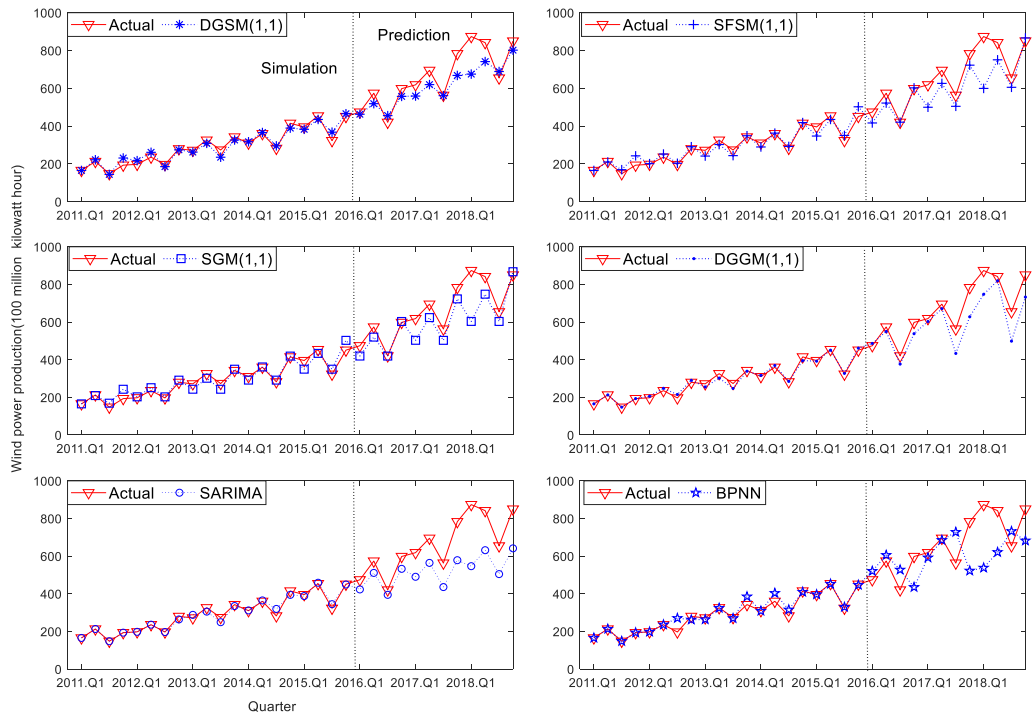

Fig 1. The split sample experiment for predicting China's wind power production by six competing models

Table 3 The forecasted results (100 million kilowatt hours), APE (%), MAPE (%), and RMSE of six competing models for quarterly wind power production

| Quarter | Actual | DGSM(1,1) | APE | SFGM(1,1) | APE | SGM(1,1) | APE | DGGM(1,1) | APE | SARIMA | APE | BPNN | APE |
|---|---|---|---|---|---|---|---|---|---|---|---|---|---|
| 2016.Q1 | 474.2 | 462.76 | 2.41 | 416.62 | 12.14 | 419.39 | 11.56 | 486.25 | 2.54 | 422.98 | 10.80 | 520.49 | 9.76 |
| 2016.Q2 | 573.8 | 518.69 | 9.60 | 521.85 | 9.05 | 520.02 | 9.37 | 550.01 | 4.15 | 510.82 | 10.98 | 605.19 | 5.47 |
| 2016.Q3 | 420.2 | 455.46 | 8.39 | 420.73 | 0.13 | 419.22 | 0.23 | 376.64 | 10.37 | 394.23 | 6.18 | 527.31 | 25.49 |
| 2016.Q4 | 598.1 | 552.30 | 6.83 | 602.27 | 0.70 | 602.98 | 0.82 | 538.17 | 10.02 | 532.45 | 10.98 | 434.82 | 27.30 |
| 2017.Q1 | 618.4 | 558.97 | 9.61 | 499.71 | 19.19 | 503.02 | 18.66 | 602.92 | 2.50 | 490.27 | 20.72 | 591.89 | 4.29 |
| 2017.Q2 | 695 | 619.54 | 10.86 | 625.92 | 9.94 | 623.73 | 10.26 | 671.37 | 3.40 | 564.01 | 18.85 | 684.76 | 1.47 |
| 2017.Q3 | 563 | 561.17 | 0.32 | 504.63 | 10.37 | 502.83 | 10.69 | 433.30 | 23.04 | 435.63 | 22.62 | 727.56 | 29.23 |
| 2017.Q4 | 783 | 668.04 | 14.68 | 722.38 | 7.74 | 723.23 | 7.63 | 627.96 | 19.80 | 578.33 | 26.14 | 522.23 | 33.30 |
| 2018.Q1 | 873.8 | 675.13 | **22.74** | 599.36 | **31.41** | 603.34 | **30.95** | 747.59 | 14.44 | 546.54 | **37.45** | 537.86 | **38.45** |
| 2018.Q2 | 841.6 | 741.31 | 11.92 | 750.74 | 10.80 | 748.12 | 11.11 | 819.52 | 2.62 | 631.45 | 24.97 | 620.73 | 26.24 |
| 2018.Q3 | 654.3 | 688.81 | 5.27 | 605.27 | 7.49 | 603.11 | 7.82 | 498.47 | **23.82** | 504.50 | 22.90 | 731.90 | 11.86 |
| 2018.Q4 | 850.1 | 801.84 | 5.68 | 866.44 | 1.92 | 867.47 | 2.04 | 732.74 | 13.81 | 641.30 | 24.56 | 682.10 | 19.76 |

| | | | | | | |
|---|---|---|---|---|---|---|
| MAPES | 6.02 | 6.83 | 6.76 | 3.07 | 4.56 | 6.42 |
| RMSES | 20.52 | 24.85 | 24.59 | 11.89 | 17.52 | 27.59 |
| MAPEP | 9.03 | 10.07 | 10.09 | 10.88 | 19.76 | 19.39 |
| RMSEP | 82.45 | 99.25 | 98.45 | 92.34 | 163.44 | 166.59 |

Note: The max APE of each model is marked in bold.

### 3.2 Quarterly PM$_{10}$ emissions

Another case for verifying the efficacy of the proposed model is to predict the emissions of PM$_{10}$ in Changzhou city in China. PM$_{10}$, technically, refers to the inhalable particles with a diameter of less than or equal to 10 microns. These particles are harmful because they can lead to serious diseases to human beings. Hence, accurate predictions may contribute to better control and treatment of PM$_{10}$ emissions. In this study, due to data availability, the quarterly PM$_{10}$ emissions from the first quarter in 2014 to the fourth quarter in 2017 are treated as the training subset and those from the first quarter in 2018 to the fourth quarter in 2019 are selected as the testing subset.

Similar to case one, six competing models are established based on the available sample data and their estimated parameters are recorded in Table 4. From this table, the mathematical properties and optimal form of these models are presented, which can help readers to repeat this experiment. Subsequently, after obtaining these parameters, the predicted results of these models can be compared in Fig. 2 and Table 5.

From Fig. 2, it can be intuitively seen that PM$_{10}$ emissions in Changzhou have a downward trend and seasonal fluctuations, which is consistent with the conclusions from the property 4 with the $\hat{\alpha} < 1$. For the fitted period, results of all six models are very close to the real observations, among which the $DGGM(1,1)$ ranks first, followed by the $DGSM(1,1)$ model. However, for the predicted period, the forecasted values of the proposed model deviate the least from the actual values, showing excellent capabilities in identifying the seasonal fluctuations in PM$_{10}$ emissions in Changzhou. For the *SARIMA* and *BPNN* models, their generated values match comparatively worse with the collected data in both fitted and predicted periods, compared with those of the grey models. Thus, the proposed model has merits in solving forecasting issues with seasonal fluctuations.

In addition, similar findings can be drawn from the perspective of accuracy levels. The $DGSM(1,1)$ model obtains the $MAPE$ of 4.33% and 6.69% in fitted and predicted periods, respectively, ranking second and first among the listed models, which is consistent with the conclusion made from $RMSE$. Although the $DGGM(1,1)$ model delivers the most accurate forecasts in the fitted period with 3.33% $MAPES$ and 4.44 $RMSES$ values, it ranks fifth of listed models in the predicted stage with 10.89% $MAPEP$ and 11.69 $RMSEP$ values, indicating worse forecasting ability. As for the $SFGM(1,1)$, $SGM(1,1)$, $SARIMA$, and $BPNN$ models, they appear to be clearly inferior to the proposed models in terms of the $MAPE$ and $RMSE$ criteria. Furthermore, with respect to the $APE$ criteria in the predicted period, the novel model still enjoys stable forecasting ability with the second smallest accuracy improvement of 20.63% $APE$ values. In conclusion, the $DGSM(1,1)$ model can cumulatively account for a dominant share of best-performing models in the presence of predicting $PM_{10}$ emissions with seasonal fluctuations.

Table 4 The estimated parameters of six competing models for quarterly $PM_{10}$ emissions

| Models | Parameters |
|---|---|
| $DGSM(1,1)$ | $\hat{P}=[0.97,134.43,111.47,84.46,124.68]$. |
| $SFGM(1,1)$ | $\hat{a}=0.028, \hat{b}=116.521, I_1=1.238, I_2=0.993, I_3=0.674, I_4=1.094$. |
| $SGM(1,1)$ | $\hat{\alpha}_s=[0.0275,116.348], f_s(1)=1.234, f_s(2)=0.994, f_s(3)=0.679, f_s(4)=1.092$. |
| $DGGM(1,1)$ | $\hat{\alpha}_1=[0.109,146.068], \hat{\alpha}_2=[0.106,113.557], \hat{\alpha}_3=[0.185,91.042], \hat{\alpha}_4=[0.148, 139.17]$ |
| $SARIMA$ | SARIMA(0,0,0)(1,1,0)$_4$, c=0, SAR(4)=-0.6924, AIC=7.183, Log L=-0.9425. |
| $BPNN$ | Learning rate=0.01, iterative number=1000, a single hidden layer, error goal=0.05, the number of neurons =13. |

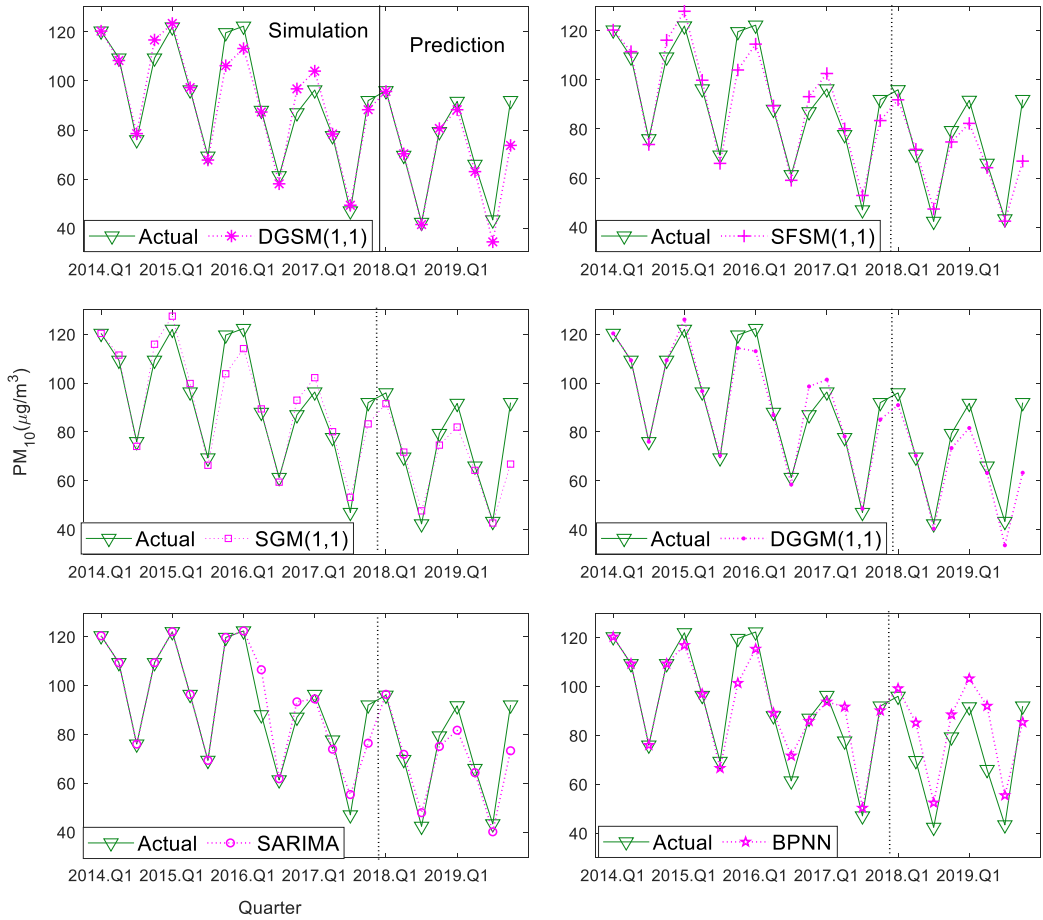

Fig 2. The split sample experiment for predicting quarterly PM$_{10}$ emissions

Table 5 The forecasted results (μg/m$^3$), APE (%), MAPE (%), and RMSE of six competing models for quarterly PM$_{10}$ emissions

| Quarter | Actual | DGSM(1,1) | APE | SFGM(1,1) | APE | SGM(1,1) | APE | DGGM(1,1) | APE | SARIMA | APE | BPNN | APE |
|---|---|---|---|---|---|---|---|---|---|---|---|---|---|
| 2018.Q1 | 96 | 95.74 | 0.27 | 91.89 | 4.28 | 91.55 | 4.64 | 90.97 | 5.24 | 96.39 | 0.40 | 99.19 | 3.32 |
| 2018.Q2 | 69.67 | 70.31 | 0.92 | 71.71 | 2.94 | 71.74 | 2.97 | 70.28 | 0.88 | 71.81 | 3.07 | 85.16 | 22.24 |
| 2018.Q3 | 42.33 | 41.47 | 2.03 | 47.36 | 11.88 | 47.68 | 12.62 | 40.40 | 4.57 | 47.92 | 13.20 | 52.40 | 23.78 |
| 2018.Q4 | 79.33 | 80.63 | 1.63 | 74.71 | 5.83 | 74.60 | 5.97 | 73.37 | 7.52 | 75.04 | 5.41 | 88.51 | 11.56 |
| 2019.Q1 | 91.67 | 88.28 | 3.69 | 82.28 | 10.24 | 82.00 | 10.55 | 81.60 | 10.98 | 81.69 | 10.88 | 103.23 | 12.61 |
| 2019.Q2 | 66 | 63.04 | 4.48 | 64.21 | 2.71 | 64.26 | 2.64 | 63.19 | 4.26 | 64.29 | 2.60 | 92.04 | **39.45** |
| 2019.Q3 | 43.33 | 34.40 | **20.63** | 42.41 | 2.13 | 42.70 | 1.46 | 33.59 | **22.49** | 40.09 | 7.49 | 55.41 | 27.87 |
| 2019.Q4 | 92 | 73.73 | 19.86 | 66.90 | **27.29** | 66.82 | **27.37** | 63.28 | 31.21 | 73.28 | **20.35** | 85.37 | 7.21 |
| MAPES |  |  | 4.33 |  | 5.51 |  | 5.40 |  | 3.33 |  | 10.07 |  | 6.58 |

| | | | | | | |
|---|---|---|---|---|---|---|
| RMSES | 5.68 | 6.22 | 6.21 | 4.44 | 10.09 | 7.87 |
| MAPEP | 6.69 | 8.41 | 8.53 | 10.89 | 7.92 | 18.51 |
| RMSEP | 7.39 | 9.94 | 10.04 | 11.69 | 8.05 | 13.40 |

### 3.3 Monthly natural gas consumption

In order to further demonstrate the efficacy and generalizability of the new model, monthly natural gas consumption is employed in case three, differing from the quarterly seasonal time sequences in the previous two cases. It is worth noting that increasing the penetration of natural gas into China's energy mix is imperative for central and local governments, due to the huge pressures from enormous energy demands and serious environmental damages from fossil fuels [3]. Thus, increasing prediction accuracy for natural gas consumption can allow governments to obtain more valuable information for formulating appropriate policies and plans. In this case, 120 observations covering from January 2009 to December 2018 are collected, which are divided into two groups, namely training data set (2009. M1-2017. M12) and testing data set (2018. M1-2018. M12).

After model calibrations of the six models, their estimated parameters are presented in Table 6. Detailed explanations of these parameters are similar to those in cases one and two and the optimal form of each model can also be formulated. Subsequently, the predicted values will be listed in Fig.3 and Table 7.

From Fig. 3, we can conclude that the predicted curve of the proposed model is much closer to the real values in the predicted period, compared with others, intuitively reflecting its excellent capability in capturing the seasonal characteristics of natural gas consumption. In addition, from the perspective of accuracy levels, the same findings can also be verified again. Specifically, the new model enjoys the third higher fitting accuracy with $MAPES$ and $RMSES$ of 7.29% and 11.25, following the $SGM(1,1)$ and $DGGM(1,1)$ models. Furthermore, in the predicted period, this proposed model obtains the lowest $MAPEP$ and $RMSEP$ values which correspond to 5.56% and 16.24, respectively, and the maximum $APE$ value is 12.66, exhibiting the excellent prediction ability. Therefore, for this and many other reasons, the $DGSM(1,1)$ model is highly recommended to use in future applications.

Table 6 The estimated parameters of six competing models for monthly natural gas consumption

| Models | Parameters |
|---|---|
| $DGSM(1,1)$ | $\hat{P} = [1.0089, 107.61, 85.14, 88.36, 70.27, 67.41, 72.63, 75.09, 75.36, 73.33, 72.65, 83.64, 98.40]\cdot$ |
| $SFGM(1,1)$ | $\hat{a} = -0.0088, \hat{b} = 80.8, I_1 = 1.108, I_2 = 0.99, I_3 = 1.03, I_4 = 0.91, I_5 = 0.93, I_6 = 0.9, I_7 = 0.94,\cdot$ $I_8 = 0.97, I_9 = 0.99, I_{10} = 0.97, I_{11} = 1.07, I_{12} = 1.16$ |
| $SGM(1,1)$ | $\hat{\alpha}_s = [-0.0088, 80.9], f_s(1) = 1.13, f_s(2) = 1, f_s(3) = 1.03, f_s(4) = 0.9, f_s(5) = 0.89, f_s(6) = 0.94,$ $f_s(7) = 0.96, f_s(8) = 0.97, f_s(9) = 0.97, f_s(10) = 0.97, f_s(11) = 1.06, f_s(12) = 1.18$ |
| $DGGM(1,1)$ | $\hat{\alpha}_1 = [-0.12, 86.81], \hat{\alpha}_2 = [-0.101, 84.28], \hat{\alpha}_3 = [-0.104, 85.39], \hat{\alpha}_4 = [-0.104, 74.77],$ $\hat{\alpha}_5 = [-0.089, 80.33], \hat{\alpha}_6 = [-0.102, 77.34], \hat{\alpha}_7 = [-0.085, 81.97], \hat{\alpha}_8 = [-0.1, 89.48],$ $\hat{\alpha}_9 = [0.10, 80.97], \hat{\alpha}_{10} = [0.10, 83.44], \hat{\alpha}_{11} = [0.0977, 90.27], \hat{\alpha}_{12} = [0.109, 100.03]$ |
| $SARIMA$ | SARIMA(1,0,1)(1,1,0)$_{12}$, c=0.13, AR(1)=0.94, MA(1)=-0.77, SAR(12)=-0.45, AIC=-1.56, Log L=79.86. |
| $BPNN$ | Learning rate=0.01, iterative number=1000, a single hidden layer, error goal=0.05, the number of neurons =13. |

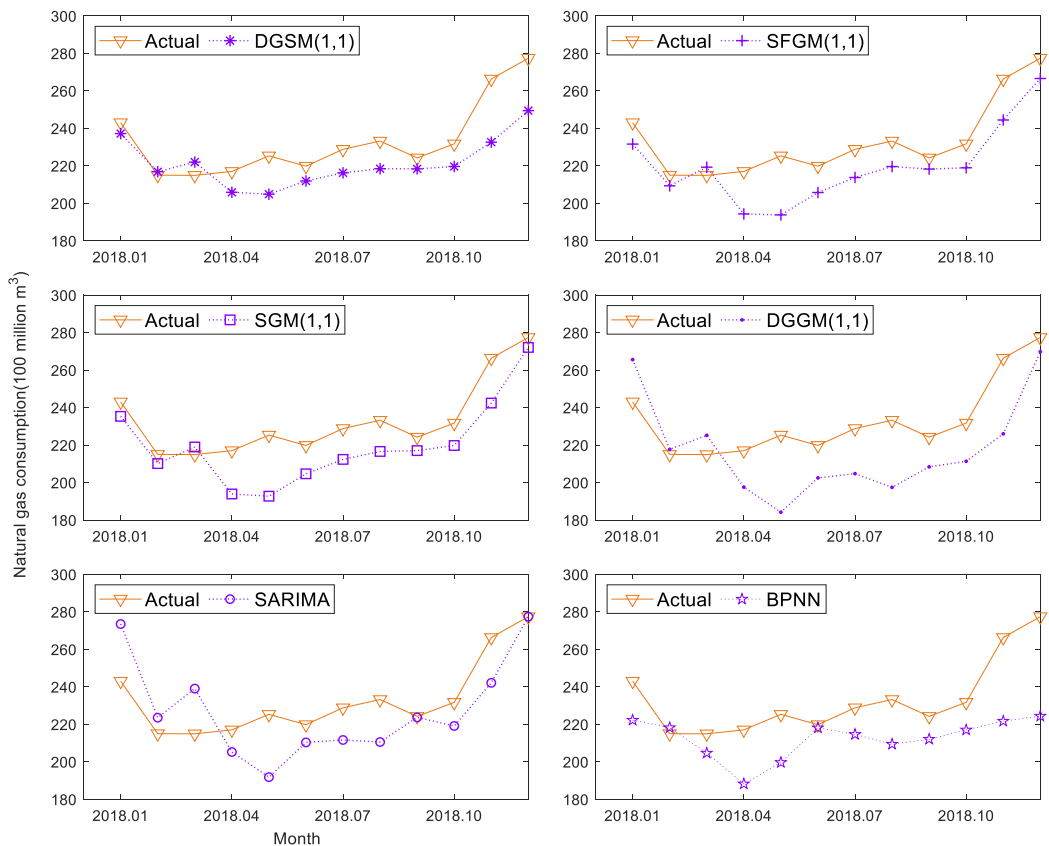

Fig 3. The predicted results of monthly natural gas consumption by six competing models

Table 7 The forecasted results (100 million m³), APE (%), MAPE (%), and RMSE of six competing models for monthly natural gas consumption

| Month | Actual | DGSM(1,1) | APE | SFGM(1,1) | APE | SGM(1,1) | APE | DGGM(1,1) | APE | SARIMA | APE | BPNN | APE |
|---|---|---|---|---|---|---|---|---|---|---|---|---|---|
| 2018.01 | 243 | 237.18 | 2.40 | 231.58 | 4.70 | 235.30 | 3.17 | 265.58 | 9.29 | 273.48 | 12.54 | 222.18 | 8.57 |
| 2018.02 | 215 | 216.82 | 0.85 | 209.32 | 2.64 | 210.10 | 2.28 | 217.67 | 1.24 | 223.53 | 3.97 | 218.19 | 1.48 |
| 2018.03 | 214.9 | 221.98 | 3.29 | 219.26 | 2.03 | 218.97 | 1.89 | 225.17 | 4.78 | 239.04 | 11.23 | 204.61 | 4.79 |
| 2018.04 | 217 | 205.87 | 5.13 | 194.31 | 10.45 | 193.88 | 10.65 | 197.54 | 8.97 | 205.22 | 5.43 | 188.18 | 13.28 |
| 2018.05 | 225.3 | 204.85 | 9.08 | 193.80 | **13.98** | 192.71 | **14.47** | 184.18 | **18.25** | 191.79 | **14.87** | 199.61 | 11.40 |
| 2018.06 | 219.8 | 211.89 | 3.60 | 205.75 | 6.39 | 204.65 | 6.89 | 202.47 | 7.89 | 210.31 | 4.32 | 218.10 | 0.77 |
| 2018.07 | 228.8 | 216.25 | 5.49 | 213.76 | 6.57 | 212.34 | 7.19 | 204.73 | 10.52 | 211.62 | 7.51 | 214.60 | 6.21 |
| 2018.08 | 233.2 | 218.45 | 6.33 | 219.60 | 5.83 | 216.58 | 7.13 | 197.43 | 15.34 | 210.54 | 9.72 | 209.37 | 10.22 |
| 2018.09 | 224.2 | 218.37 | 2.60 | 218.26 | 2.65 | 217.03 | 3.20 | 208.34 | 7.07 | 223.67 | 0.23 | 212.03 | 5.43 |
| 2018.10 | 231.7 | 219.64 | 5.20 | 218.94 | 5.51 | 219.75 | 5.16 | 211.34 | 8.79 | 219.08 | 5.44 | 216.97 | 6.36 |
| 2018.11 | 266.3 | 232.59 | **12.66** | 244.44 | 8.21 | 242.39 | 8.98 | 225.99 | 15.14 | 242.12 | 9.08 | 221.66 | 16.76 |
| 2018.12 | 277.4 | 249.43 | 10.08 | 266.58 | 3.90 | 272.09 | 1.91 | 269.81 | 2.74 | 277.40 | 0.00 | 224.34 | **19.13** |
| MAPES | | | 7.29 | | 7.36 | | 7.24 | | 5.02 | | 7.35 | | 7.78 |
| RMSES | | | 11.25 | | 11.53 | | 11.29 | | 8.47 | | 13.12 | | 12.60 |
| MAPEP | | | 5.56 | | 6.07 | | 6.08 | | 9.17 | | 7.03 | | 8.70 |
| RMSEP | | | 16.24 | | 16.05 | | 16.50 | | 24.50 | | 19.34 | | 25.80 |

### 3.4 Summary of the above three studies

Following the previous discussions of the three studies, several conclusions can be further summarized as follows:

(1) The proposed $DGSM(1,1)$ model can efficiently identify seasonal fluctuations in the diverse cases and its quantitative improvements over forecasts are even more pronounced than those of the existing prevalent seasonal models, referring to $SFGM(1,1)$, $SGM(1,1)$, $DGGM(1,1)$, $SARIMA$, and $BPNN$. These successful projections are strongly attributed to the incorporation of seasonal dummy variables into the conventional discrete grey model.

(2) The $DGGM(1,1)$ model achieves the best performance in the fitted period and the worst

in the predicted period among the listed grey models, which means that this model has poor performance in future estimations. As for the $SFGM(1,1)$ and $SGM(1,1)$ models, they model the seasonal sequences in a similar way that they preprocess the original series to remove seasonal fluctuations by using seasonal factors. Small differences lie in the approaches to calculating the factors. Thus, the forecasted values by these two grey models are very close in the above three cases, which can be verified in Tables 3, 5, and 7. Moreover, these two models achieve better overall performance than the $DGGM(1,1)$ model, thereby being appropriate for the alternative methodologies to model seasonal sequences.

(3) For the $SARIMA$ and $BPNN$ models, they both can obtain high fitting accuracy for the three cases. However, they generally generate much larger errors than the grey models. Reasons for such situations mainly refer to their inherent requirements for a large amount of data for model training. Without sufficient observations, they may produce unacceptable errors in real experiments, which can be validated from the comparisons between cases one and three, or cases two and three.

In conclusion, the proposed model does significantly improve forecasting performance in the presence of seasonal sequences and has high reliability and generalizability in confronting monthly or quarterly series with various sample sizes. Subsequently, the adaptability of the novel model will be further tested in the following subsection.

**3.5 Robustness of the proposed model for different sequence lengths**

In the above three cases, the length of available data, remarked as $n$, all satisfy $n \bmod s = 0$, meaning that the observations have complete cycles. However, due to data availability, real collected data may not have full periods, namely $n \bmod s \neq 0$. Under such situations, the efficacy of the proposed model is waiting to be further demonstrated. In addition, it is uncertain whether the proposed model can still make accurate forecasts, confronting sequences with different lengths. To this end, in-depth discussions are necessary for checking the robustness of the proposed model.

We take the natural gas consumption (120 observations from January 2009 to December 2018) as an example for verifying the robustness of the proposed model in comparison with other grey models. Before examinations, some important concepts about the sequence lengths required by different models should be identified. For the $SFGM(1,1)$, $SGM(1,1)$, and $DGSM(1,1)$ models, more than two complete cycles ($n \geq 2s$) are necessary for model calibrations. In other words, at least 24 observations of natural gas consumption for these above three grey models must be guaranteed. Additionally, for the $DGGM(1,1)$ model, the required sequence lengths must reach more than four cycles ($n \geq 4s$) because each separated group must have at least four data points for modeling the $GM(1,1)$ model. That is, experiments on natural gas consumption by using the $DGGM(1,1)$ model must have more than 48 observations.

Due to the similar performance of the $SFGM(1,1)$ and $SGM(1,1)$ models, explained in Section 3.4, the $SFGM(1,1)$ model will be selected to compare with the proposed model. Thus, three grey models, namely $SFGM(1,1)$, $DGGM(1,1)$, and $DGSM(1,1)$, are used to conduct experiments natural gas consumption in order to verify their robustness when modeling seasonal sequences with different lengths. The detailed experiments are designed as follows:

For the $SFGM(1,1)$ and $DGSM(1,1)$ models, the preceding 108 observations (2009. M1-2017. M12) are used to forecast the last 12 data sets (2018. M1-2018. M12). Subsequently, removing the oldest data point (2009.M1), 107 observations (2009. M2-2017. M12) are utilized to predict the last 12 values (2018. M1-2018. M12). Repeat these above operations, in the end, 24 observations (2016. M1-2017. M12) are employed to project the last 12 values (2018. M1-2018. M12).

In addition, for the $DGGM(1,1)$ model, the preceding 108 observations (2009. M1-2017. M12) are used to forecast the last 12 data sets (2018. M1-2018. M12). Subsequently, removing the oldest data point (2009. M1), 107 observations (2009. M2-2017. M12) are utilized to predict the last 12 values (2018. M1-2018. M12). Repeat these above operations, in the end, 48 observations (2014. M1-2017. M12) are employed to project the last 12 values (2018. M1-2018. M12).

By conducting these experiments, the $MAPES$ and maximum $APE$ in the predicted period are recorded, as seen in Fig. 4. It can be shown that the proposed model consistently obtains the lowest $MAPES$ values when dealing with different sequence lengths, followed by the $SFGM(1,1)$ model, while the $DGGM(1,1)$ model produces the largest $MAPES$ values. In addition, from the perspective of maximum $APE$ values, similar findings can be made that the new model still produces the overall low level of $APE$ values, even with slight fluctuations, when confronting diverse lengths of modeling series. In general, concluded from Fig. 4, the $DGSM(1,1)$ model has strong robustness and high reliability in addressing seasonal sequences with different lengths.

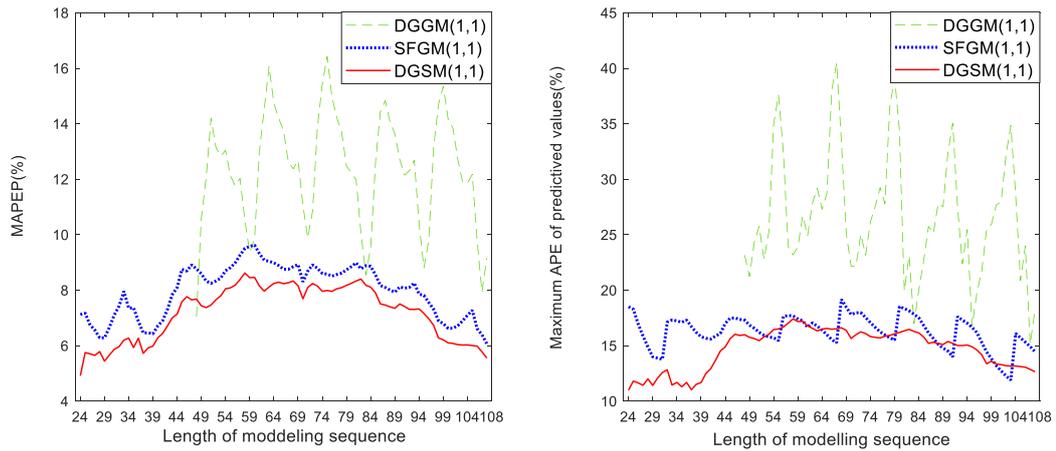

Fig.4 The MAPEP and maximum APE values of the three grey models with different sequence lengths

## 4 Conclusions and future work

Based on the characteristics of time series with seasonal fluctuations, a novel discrete grey seasonal model, i.e. $DGSM(1,1)$, is designed in order to accurately identify the seasonality. In this proposed methodology, two core innovations are included. Firstly, the discrete grey seasonal model is put forward by introducing seasonal dummy variables into the conventional discrete grey model. Moreover, detailed mathematical derivations of the parameter estimations are provided. Secondly, the properties of this new model are comprehensively analyzed in order to show the connections and differences with the conventional grey model.

For verification purposes, the newly-designed methodology is implemented to forecast quarterly wind power production, quarterly $PM_{10}$, and monthly natural gas consumption in China. The consistent conclusions from the experimental results demonstrate that the proposed model strikingly improves the forecasting precision since it can achieve better performance in both fitted and predicted periods. Moreover, this new model has been proved to be more robust and reliable than the existing grey seasonal models in the presence of seasonal sequences with different lengths.

This further confirms that the newly-designed method is a very promising tool in analyzing and predicting time series with seasonal fluctuations.

However, besides the above three cases, we believe the proposed methodology can do well in other applications, referring to forecasting seasonal time series. Moreover, this paper mainly focuses on the univariate grey model. Thus, how to extend this idea into multivariable analysis may be a good research direction.

## 5 Declaration of competing interest

The authors declared that they have no conflicts of interest in this work.

**Acknowledgments**

The authors really appreciate the time of editors and the anonymous reviewers for reviewing our paper. This work was funded by the National Natural Science Foundation of China (Grant Nos 71701024, 71901191, 71971194, 71703144) and National Social Science Foundation of China (Grant Nos 16CTJ005). The project of the philosophy and Social Sciences in Hangzhou (M20JC086).

# Reference

[1] Jelica D, Taljegard M, Thorson L, et al. Hourly electricity demand from an electric road system – A Swedish case study. Applied Energy 2018; 228(15):141-148.

[2] Venkata R, Krishna B, Kumar S R, et al. Monthly Rainfall Prediction Using Wavelet Neural Network Analysis. Water Resour Manage, 2013,27: 3697–3711.

[3] Wang Q, Li S Y, Li R R, et al. Forecasting U.S. shale gas monthly production using a hybrid ARIMA and metabolic nonlinear grey model, Energy, 2018,160 :378-387.

[4] De Felice M, Alessandri A, Catalano F. Seasonal climate forecasts for medium-term electricity demand forecasting. Applied Energy 2015; 137:435–444.

[5] Xu S, Chan H K, Zhang T. Forecasting the demand of the aviation industry using hybrid time series SARIMA-SVR approach. Transportation Research Part E: Logistics and Transportation Review 2019; 122:169-180.

[6] Proietti, T. Seasonal heteroscedasticity and trends. Journal of Forecasting 1998;17(1):1-17.

[7] Penzer J, Tripodis Y. Single-season heteroscedasticity in time series. Journal of Forecasting 2007; 26(3):189-202.

[8] Proietti, Tommaso M, Martyna M, Gianluigi. A class of periodic trend models for seasonal time series. Journal of Forecasting 2019; 38(2):106-121.

[9] Javed F, Ghim P O. Forecasting seasonal container throughput at international ports using SARIMA models. Maritime Economics & Logistics 2016; 21: 1-18.

[10] Tsui W H K, Ozer Balli H, Gilbey A, et al. Forecasting of Hong Kong airport's passenger throughput. Tourism Management 2014; 42:62-76.

[11] Miloš M, Libor Š, Vlastimil M, et al. Sarima modeling approach for railway passenger flow forecasting. Transport 2018; 33(5): 1113–1120.

[12] Gökhan S. The Efficacy of SARIMA Models for Forecasting Inflation Rates in Developing Countries: The Case for Turkey. International Research Journal of Finance and Economics 2011; 62: 1450-2887.

[13] Mwanga D, Ong'ala J, Orwa G. Modeling Sugarcane Yields in the Kenya Sugar Industry: A SARIMA Model Forecasting Approach. International Journal of Statistics and Applications 2017; 7(6): 280-288.

[14] Danandeh M, Kahya A, Sahin, et al. Successive-station monthly streamflow prediction using different artificial neural network algorithms. International Journal of Environmental Science and Technology 2015; 12(7): 2191-2200.

[15] Khwaja A S, Anpalagan A, Naeem M, et al. Joint bagged-boosted artificial neural networks: Using ensemble machine learning to improve short-term electricity load forecasting. Electric Power Systems Research 2020;179: 106080.

[16] Li S W, Chen T, Wang L, et al. Effective tourist volume forecasting supported by PCA and improved BPNN using Baidu index. Tourism Management 2018; 68:116-126.

[17] Omer F B, Beyzanur C E, Ekrem T, et al. Using machine learning tools for forecasting natural gas consumption in the province of Istanbul. Energy Economics 2019; 80: 937-949.

[18] Xie G, Wang SY, Zhao YX, et al. Hybrid approaches based on LSSVR model for container throughput forecasting: A comparative study. Applied Soft Computing 2013; 13: 2232-2241.

[19] Wei B L, Xie N M, Yang L. Understanding cumulative sum operator in grey prediction model with integral matching. Communications in Nonlinear Science and Numerical Simulation 2020: 105076.

[20] Deng J L. Control problem of grey systems. 1982; 1(5): 288-294.

[21] Deng J L. Basic methods of grey system. Wuhan: Central-China Science and Technology Press 2005.

[22] Lee Y S, Tong L. Forecasting energy consumption using a grey model improved by incorporating genetic programming. Energy Conversion and Management 2011; 52(1):147-152.

[23] Ye J, Dang Y, Li B. Grey-Markov prediction model based on background value optimization and central-point triangular whitenization weight function. Communications in Nonlinear Science and Numerical Simulation 2018; 54: 320-330.

[24] Wu L F, Liu SF, Yao L, et al. The effect of sample size on the grey system model. Applied Mathematical Modelling 2013; 37(9): 6577-6583.

[25] Ma X, Hu YS, Liu ZB. A novel kernel regularized nonhomogeneous grey model and its applications. Communications in Nonlinear Science and Numerical Simulation 2017; 48:51-62.

[26] Coskun H, Huseyin AE. Forecasting the annual electricity consumption of Turkey using an optimized grey model. Energy 2014; 70:165-171.

[27] Ding S, Dang YG, Li XM, et al. Forecasting Chinese $CO_2$ emissions from fuel combustion using a novel grey multivariable model. Journal of Cleaner Production 2017; 162:1527-1538.

[28] Zeng B, Duan HM, Bai Y, et al. Forecasting the output of shale gas in China using an unbiased grey model and weakening buffer operator. Energy 2018; 151:238-249.

[29] Li C, Yang Y, Liu S. Comparative analysis of properties of weakening buffer operators in time series prediction models[J]. Communications in Nonlinear Science and Numerical Simulation 2019, 68: 257-285.

[30] Ding S. A novel discrete grey multivariable model and its application in forecasting the output value of China's high-tech industries. Computers & Industrial Engineering, 2019; 127: 749-760.

[31] Wang J, Ma X, Wu J, et al. Optimization models based on GM (1,1) and seasonal fluctuation for electricity demand forecasting. International Journal of Electrical Power & Energy Systems 2012; 43(1):109-117.

[32] Wang Z X, Li Q, Pei L L. A seasonal GM(1,1) model for forecasting the electricity consumption of the primary economic sectors. Energy 2018; 154:522-534.

[33] Wang ZX, Li Q, Pei LL. Grey forecasting method of quarterly hydropower production in China based on a data grouping approach. Applied Mathematical Modelling 2017; 51: 302-316.

[34] Xie N M, Liu S F. Discrete grey forecasting model and its optimization. Applied Mathematical Modelling, 2009, 33(2):1173-1186.

[35] Mombeni H A, Rezaei S , Nadarajah S , et al. Estimation of Water Demand in Iran Based on SARIMA Models. Environmental Modeling & Assessment, 2013, 18(5):559-565.

[36] Ding S, Xu N, Ye J, et al. Estimating Chinese energy-related CO2 emissions by employing a novel discrete grey prediction model. Journal of Cleaner Production 2020; 120793. https://doi.org/10.1016/j.jclepro.2020.120793

[37] Hu Y C. Electricity consumption prediction using a neural-network-based grey forecasting approach. Journal of Operational Research Society 2017; 68: 1259-1264.